\documentclass[journal,twoside,web]{ieeecolor}
\usepackage{tmi}
\usepackage{cite}
\usepackage{amsmath,amssymb,amsfonts}
\usepackage{algorithm}
\usepackage{algpseudocode}
\usepackage{graphicx}
\usepackage{textcomp}
\usepackage{multirow}
\usepackage{hyperref}
\usepackage[table]{xcolor}

\makeatletter
\newcommand\fs@norules{\def\@fs@cfont{\bfseries}\let\@fs@capt\floatc@ruled
  \def\@fs@pre{}%
  \def\@fs@post{}%
  \def\@fs@mid{\kern3pt}%
  \let\@fs@iftopcapt\iftrue}
\makeatother
\floatstyle{norules}
\def\BibTeX{{\rm B\kern-.05em{\sc i\kern-.025em b}\kern-.08em
    T\kern-.1667em\lower.7ex\hbox{E}\kern-.125emX}}
\begin{document}
\title{Less is More: Unsupervised Mask-guided Annotated CT Image Synthesis with Minimum Manual Segmentations}
\author{Xiaodan Xing, Giorgos Papanastasiou, Simon Walsh* and Guang Yang*,~\IEEEmembership{Senior Member, IEEE}
\thanks{*Joint senior authors}
\thanks{X. Xing, S. Walsh and G. Yang are with the National Heart and Lung Institute, Imperial College London, UK (send correspondence to x.xing@imperial.ac.uk and g.yang@imperial.ac.uk).}
\thanks{S. Walsh and G. Yang are also with the Royal Brompton Hospital, UK.}
\thanks{Giorgos Papanastasiou is with the Centre for Cardiovascular Science, University of Edinburgh, UK and the School of Computer Science and Electronic Engineering, University of Essex, UK.}
\thanks{This study was supported in part by the BHF (TG/18/5/34111, PG/16/78/32402), the ERC IMI (101005122), the H2020 (952172), the MRC (MC/PC/21013), the Royal Society (IEC/NSFC/211235), the NVIDIA Academic Hardware Grant Program, the SABER project supported by Boehringer Ingelheim Ltd, NIHR Imperial Biomedical Research Centre (RDA01), and the UKRI Future Leaders Fellowship (MR/V023799/1).}}
\maketitle

\begin{abstract}
As a pragmatic data augmentation tool, data synthesis has generally returned dividends in performance for deep learning based medical image analysis. However, generating corresponding segmentation masks for synthetic medical images is laborious and subjective. To obtain paired synthetic medical images and segmentations, conditional generative models that use segmentation masks as synthesis conditions were proposed. However, these segmentation mask-conditioned generative models still relied on large, varied, and labeled training datasets, and they could only provide limited constraints on human anatomical structures, leading to unrealistic image features. Moreover, the invariant pixel-level conditions could reduce the variety of synthetic lesions and thus reduce the efficacy of data augmentation. To address these issues, in this work, we propose a novel strategy for medical image synthesis, namely Unsupervised Mask (UM)-guided synthesis, to obtain both synthetic images and segmentations using limited manual segmentation labels. We first develop a superpixel based algorithm to generate unsupervised structural guidance and then design a conditional generative model to synthesize images and annotations simultaneously from those unsupervised masks in a semi-supervised multi-task setting. In addition, we devise a multi-scale multi-task Fréchet Inception Distance (MM-FID) and multi-scale multi-task standard deviation (MM-STD) to harness both fidelity and variety evaluations of synthetic CT images. With multiple analyses on different scales, we could produce stable image quality measurements with high reproducibility. Compared with the segmentation mask guided synthesis, our UM-guided synthesis provided high-quality synthetic images with significantly higher fidelity, variety, and utility ($p<0.05$ by Wilcoxon Signed Ranked test).

\end{abstract}


\section{Introduction}
\label{sec:introduction}
\IEEEPARstart{S}{YNTHETIC} medical images generated by Generative Adversarial Networks (GANs) can provide efficient and promising data augmentation for medical AI algorithms  \cite{frid2018synthetic,han2019synthesizing}. These synthetic images can improve the quantity and variety of training datasets and enhance the performance of downstream AI model tasks, such as segmentation and classification.
 \begin{figure*}[t]
\centerline{\includegraphics[width=17cm]{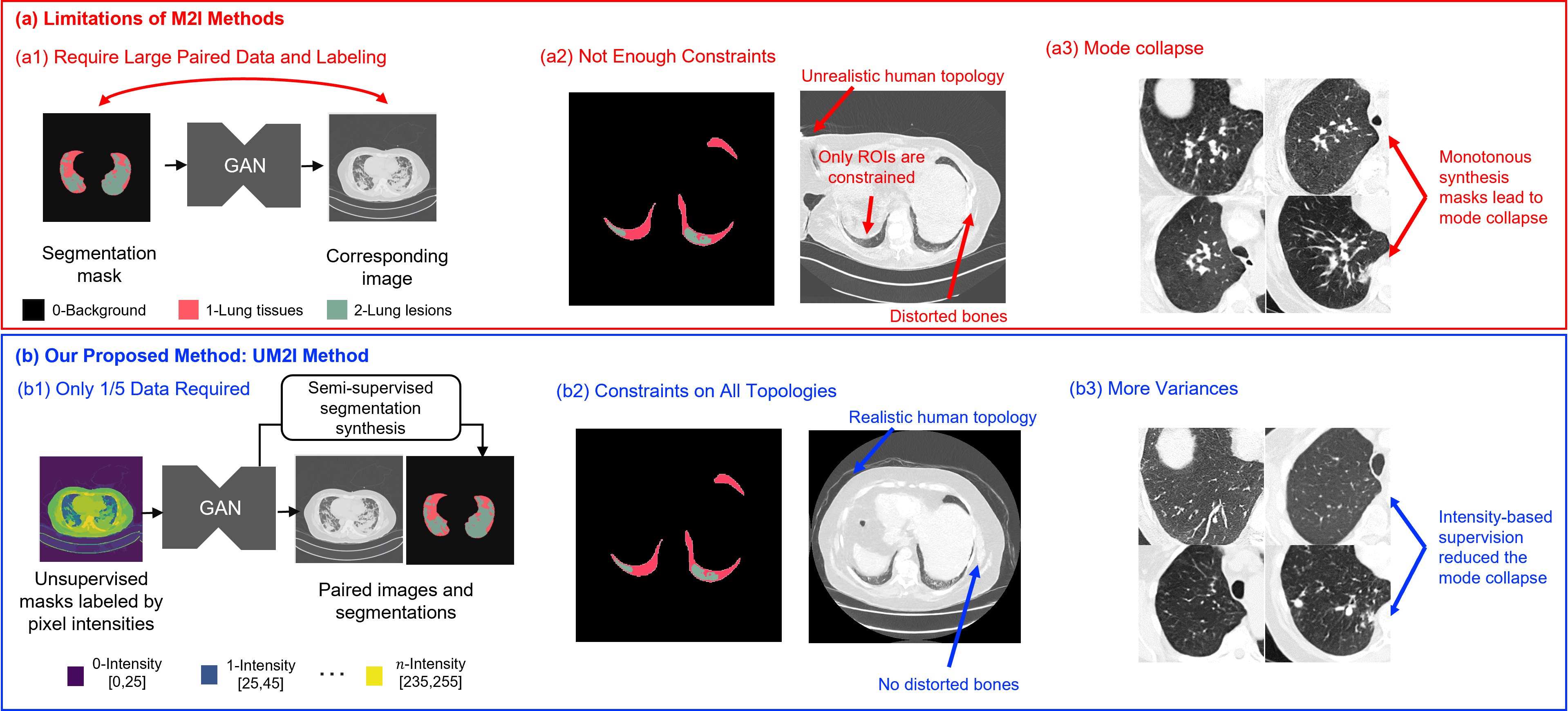}}
\caption{Limitations of M2I strategies (a) in the red panel and our proposed solutions (b) in the blue panel. We identified three major problems with M2I algorithms including the large training data requirements (a1), limited constraints (a2), and confined variety (a3) that could limit the power of data augmentation. In this paper, we proposed a UM2I synthesis strategy. We used a multi-task strategy to synthesize images and corresponding segmentations from unsupervised masks, which allows segmentation synthesis using only a limited number of manual labeling (b1). In addition, our intensity based unsupervised masks provide constraints on human anatomies that are not specifically labeled in segmentation masks, producing realistic human anatomies synthesis (b2). Last but not least, unlike the segmentation guided synthesis only provide patient invariant conditions on all regions, our UM-guided synthesis provide additional intensity distribution as guidance information for all tissues and can model the full spectrum of the CT HU distributions \textcolor{black}{among all test cases (b3).} }
\label{fig:teaser}
\end{figure*}

However, obtaining the annotations for synthetic images remains challenging. Existing GAN-based medical image synthesis algorithms \cite{goodfellow2014generative,karras2019stylegan,karras2020stylegan2} do not synthesize the corresponding segmentation annotations for synthetic images intrinsically. It is also impossible to manually label all synthetic images (e.g., manual segmentation) as it is tedious and subjective. For cross-modality synthesis that generates images from different modalities, synthetic images may inherit the segmentation masks from their corresponding inputs \cite{oulbacha2020mri} if the deformation among different imaging modalities can be retrieved. But for intra-modality synthesis, the lack of synthetic labels limits the application of synthetic medical images. To address this issue, some researchers proposed a conditional synthesis strategy \cite{pandey2020image,liang2022sketch} in intra-modality synthesis, where the images were synthesized from segmentation masks. By generating a large amount of segmentation masks, synthetic images with realistic human anatomies or pathologies can be generated. This can be a powerful data augmentation strategy as it can generate paired images and segmentations and thus, can bring breakthroughs in general AI modeling downstream tasks such as segmentation and classification. In this manuscript, we refer to this synthesis process as the Mask-to-Image (M2I) synthesis, because the method is conditioned on manual segmentation masks.

Although the M2I synthesis can be promising, three major problems hinder the real-world applications of this strategy. First, a well-performed M2I network requires large-scale training data \cite{pandey2020image}, i.e., a large paired image and segmentation mask dataset is required, which contradicts the motivation of solving the scarcity of data annotations (Fig. \ref{fig:teaser} (a1)). Second, M2I synthesis typically receives conditions only on Regions of Interest (ROI), setting no or limited guidance on other human anatomical structures and pathological features, that may lead to unrealistic hallucinated structures overall. An example can be found in Fig. \ref{fig:teaser} (a2): in a lung lesion segmentation task, the segmentation masks fail to provide enough constraints on the human body and bones, leading to unrealistic anatomy representations. Third, the M2I synthesis uses subject-invariant conditions on regions among all cases, so the M2I model can easily collapse due to dataset-specific biases. For example, COVID-related lesions manifested in CT images include ground glass opacities (GGO, which are hazy and increase lung opacity without obscuration of the underlying parenchymal vessels) and consolidation (an increase in lung opacity that does obscure the vessels). In the COVID-related lesion synthesis task, M2I models synthesized these various lesions with invariant labels and thus failed to cover the histograms of all COVID-related lesions (Fig. \ref{fig:teaser} (a3)), especially for lesions with HU values ranging between -300 and 0, which are the common values for the GGO pattern \cite{roth2021rapid}.

\begin{figure*}[t]
\centerline{\includegraphics[width=17cm]{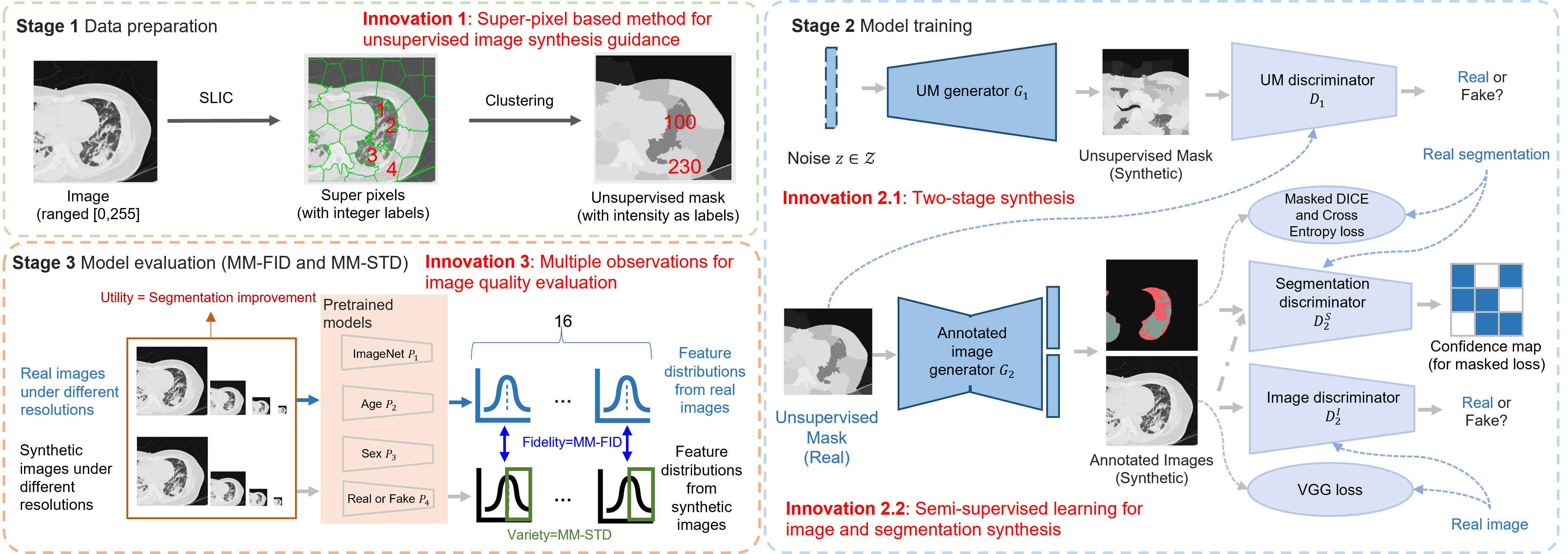}}
\caption{Overview of the proposed method. We used a superpixel based method to generate a structural guidance for image synthesis and image editing in an unsupervised manner (stage 1). During the training stage, we used real images and structural maps derived from real images to train the guidance synthesizer and image synthesizer (stage 2). At the inference stage, our method could generate realistic images with annotations from a vector of noises. We also evaluated the quality of synthetic images with multi-scale features from multiple feature extractors and provided statistical analysis for the image quality evaluation (stage 3).}
\label{fig:flowchart}
\end{figure*}

To address these three major obstacles in existing M2I models, we propose a novel method named Unsupervised Mask-to-Image (UM2I) in conditional medical image synthesis, which describes the development of an innovative unsupervised mask generation process and is subsequently incorporated into a semi-supervised learning algorithm.\textcolor{black}{The overview of our proposed UM2I synthesis strategy is shown in Fig. \ref{fig:flowchart}.} Compared to conventional M2I methods, our UM2I can obtain accurate segmentation masks with a limited number of human labeling (Fig. \ref{fig:teaser} (b1)). Essentially, our UM2I synthesis model reduces the complexity of the segmentation mask generation, as the unsupervised masks generated, provide a preliminary but effective clustering for all pixels. With this preliminary clustering, only a small fraction of labeled images is required to generate accurate segmentation masks for synthetic images. In addition, the unsupervised masks are easy to generate and provide adequate constraints globally within each image for human anatomical structures, even for those non-masked anatomical structures (Fig. \ref{fig:teaser} (b2)). Last but not least, our unsupervised masks can simulate the signal intensity distributions of real images. These intensity distributions provide varied conditions for different images and can increase the variety of synthetic images and reduce mode collapse problems (Fig. \ref{fig:teaser} (b3)). 

The method proposed in this paper is an extension of our conference submission \cite{xing2022cs}, in which a proof of concept study on UM2I was first proposed. Although we showed promising results, we identified several limitations in the original architecture during large-scale implementation and evaluation. Thus, compared to our previous implementation, this extension work has several major contributions and strengths that enhance further the performance and generalizability of our initial proof of concept study:

1. In this study, we propose a novel unsupervised mask generation method based on superpixels, which is more stable and editable compared to the previously proposed CNN based unsupervised mask generation. Compared to the original conference submission, the number of clusters on unsupervised masks generated by our new method is controllable for all images, and we can accurately adjust the extent of details for the unsupervised structural guidance by tuning the parameters of the generation method.  

2. We also propose an innovative multi-task learning based semi-supervised strategy for the synthesis of segmentation masks. We discover that the image synthesis task and the image segmentation task share commonalities, i.e., both tasks require rich semantic information to achieve a reasonable performance. Thus, we combined these two tasks in a multi-task setting. To reduce the number of labels required for training, we introduce a semi-supervised strategy in the multi-task setting. Compared to our conference submission, this multi-task learning based segmentation synthesis method not only reduces the computational costs for segmentation synthesis but also improves the segmentation performance with assistance during the synthesis task.

3. We develop an extensive multi-scale multi-task image quality evaluation for synthetic images. To capture the distribution of images and be consistent with human perception, quality evaluations are mostly operated in the feature space. The conventional evaluation method only extracts perceptual features from one model pretrained on ImageNet, which has different data distributions compared to medical images and might lead to observation bias. In this paper, we propose Multi-scale multi-task Fréchet Inception Distance (MM-FID) and multi-scale multi-task Standard Deviation (MM-STD), which incorporate different feature extractors under different resolutions to extract more representative features. MM-FID and MM-STD enable multiple observations of dataset distributions and can produce stable and reproducible analysis results for image fidelity evaluation.

\section{Related Studies}
\textcolor{black}{Medical images with complete annotations are difficult to obtain, which can compromise the performance of AI-based medical image segmentation and classification. To address this data insufficiency issue, solutions have been proposed, including semi-supervised learning algorithms and transfer-learning algorithms.}

\textcolor{black}{Semi-supervised segmentation algorithms, such as pseudo-label generation \cite{hung2018semiadv}, although can reduce the amount of data used in deep learning algorithms, still require a labeled dataset whose size matches with the unlabeled dataset. Besides, it does not tackle image insufficiency problem. Transfer learning algorithms\cite{yosinski2014transferable,tzeng2017adversarial} utilize knowledge from a large labeled dataset from a related task. However, these algorithms often require manual adjustments to masks or pre-labeling, which do not alleviate the human workload. In addition, domain adaptation algorithms may not be suitable when the target task lies in a domain that is far from the source task and there is currently a lack of consensus and defined methodologies that can efficiently perform transfer learning across disparate tasks.}

To address these issues, synthesizing images conditioned on segmentation masks has been proposed \cite{pandey2020image,liang2022sketch,mok2018learning}. These algorithms synthesized images in an M2I manner. For example, Mok et al. \cite{mok2018learning} used manual segmentation masks to generate augmented brain tumor images with different modalities.  Liang et al. \cite{liang2022sketch} used a combination of background labeling and manual delineation of ROIs to generate editable input masks. However, these methods relied heavily on the availability of numerous segmentation masks, contradicting the motivation of solving the scarcity of data annotations. Moreover, these annotations lack global constraints on surrounding tissues, and can only provide subject-invariant anatomical conditions, thus reducing the variety of synthetic images. This lack of variety can sabotage the efficiency of these synthetic images for pragmatic data augmentation.  

To increase the quantity and variety of the input masks, Pandey et al. \cite{pandey2020image} proposed to use an additional GAN model to generate input segmentation masks. This additional GAN model received a vector of noise as input and generated an unlimited number of segmentation masks. The inferences then followed a Vector-to-Mask-to-Image (V2M2I) pipeline. V2M2I based methods could increase the variety of synthetic images to a great extent, but these methods were still label-craving and failed to provide proper constraints to account for different human anatomical structures. The lack of constraints also increased the requirement for large training datasets, exacerbating the synthesis performance in data labeling deficiency scenarios.

 Recently, diffusion models \cite{rombach2022high} have been proposed to improve both the quality and variety of conditional image synthesis. However, the segmentation-guided diffusion models are still data-craving and requires large scale of explicit supervision. In addition, the training time and required GPU resources for diffusion models can be demanding, e.g., stable diffusion requires 150,000 hours of training on an NVIDIA A100-40GB \cite{rombach2022high}. Therefore, advanced GAN based models are still prevalent in various medical image synthesis tasks. 
 
In addition to the conditional synthesis based solutions, recently, Zhang et al. proposed datasetGAN \cite{zhang2021datasetgan} to synthesize annotated images from vectors of noises. They first trained a StyleGAN \cite{karras2019stylegan} to produce synthetic images. Then they selected a handful of these synthetic images and retrieved the features of these synthetic images from the StyleGAN. They labeled these synthetic images and trained a pixel-wise classification network with the features retrieved as inputs and manual labeling as outputs. Once both networks were trained successfully, a vector of noise could be used to first generate images, then to obtain annotations with the pixel-wise classification network. The datasetGAN reduced the requirement of a large dataset, but the pixel-wise classification required complex modelling steps, consumed large GPU memories for high-resolution images and increased the training time to a great extent.
 
% In addition to the training pipeline, to improve the synthesis performance of GAN models, novel training strategies \cite{karras2017progressive,brock2018large,wang2018pix2pixhd}, novel network structures \cite{huang2017adain,karras2019stylegan} and loss functions \cite{johnson2016perceptual} have been proposed. 

\section{Materials and Methods}
\subsection{Stage 1: Data Preparation}
In our previous work \cite{xing2022cs}, we used a CNN based network to obtain the unsupervised guidance on synthesized images. However, we replaced the CNN component in this study with the following methodology, because we observed that the CNN could not guarantee a stable number of clusters in the unsupervised mask generation process. This instability can decrease the controllability and reproducibility of our synthesis model. 
 \begin{algorithm}[h]
 \caption{Pseudocode of generating unsupervised masks}
 \label{alg:UMclustering}
 \begin{algorithmic}[1]
 \Require An normalized image $I$, the number of superpixels $M$, the intensity value gap $t\in(0,255)$
 \Ensure An unsupervised mask $H$
  \State $S$ $\leftarrow$ GetSuperPixels($I$,$M$)   \Comment{Get $M$ superpixels from $I$}
  \For {$m = 1$ to $M$} \Comment{Assign the mean intensity values to each superpixel}
  \State x,y=Where(S=m)
  \State S[x,y]=mean(I[x,y])
  \EndFor
  \State $H\leftarrow int(S/t)\times t$ \Comment{Cluster superpixels according to their intensity values}
 \State \textbf{return} $S$
 \end{algorithmic}
 \end{algorithm}
In this study, we proposed a superpixel guided algorithm to obtain the structural map in an unsupervised manner. We adopted the concept of "superpixels" to generate spatially over-segmented regions, which were widely used in medical image analysis applications with proven efficacy and reproducibility. superpixels are clusters of pixels that share similar properties, such as intensity and proximity \cite{achanta2010slic}. By grouping perceptually similar pixels into superpixels, regions with different clinical contexts can be demarcated. More specifically, we first used a simple linear iterative clustering (SLIC) \cite{achanta2010slic} algorithm to generate $M$ superpixels from the input image. SLIC measures the color similarity and proximity of all pixels and generates superpixels by clustering. It is of note that many other superpixel algorithms exist that vary from graph-based algorithms \cite{ren2003superpixel} to patch-based algorithms \cite{drucker2009pathfider}. We chose SLIC for our development because (1) it is relatively more efficient, especially for high-resolution images; (2) it is compatible with grey-scale images, which can be applied to most medical images, and (3) it has only two parameters to be tuned, and therefore enhance the reproducibility of our proposed synthesis framework.  

We further clustered the superpixels into super-clusters to better represent the anatomical information according to their intensity values. For all images normalized into $[0, 255]$, the pseudocode for generating unsupervised masks from superpixels is presented in Pseudocode \ref{alg:UMclustering}. Two parameters were required to generate the unsupervised guidance: the number of superpixels $M$ and the clustering threshold $t$. These two parameters are not independent of each other: the threshold value $t$ determines the level of details provided in the unsupervised masks (see ablation experiment about how optimum values for these two parameters were selected).

\subsection{Stage 2: Model Training}
\subsubsection{Unsupervised Mask Generation }
Conventional algorithms derived the masks from real images, thus limiting the variety of synthetic images. Hence, we proposed an unsupervised mask synthesis network $G_1$ to generate a large amount of unsupervised masks. In theory, with $G_1$, unlimited numbers of unsupervised masks could be generated from vectors of noises, which could subsequently improve the image synthesis variety to a great extent. 

Practically, we developed a StyleGAN \cite{karras2019stylegan} based architecture to generate unsupervised masks. The StyleGAN model is featured by a style based generator, which was used to solve the feature entanglement problem and to improve the variety of synthetic unsupervised masks. In addition, the progressive growing strategy of the StyleGAN allows a high-resolution synthesis for unsupervised masks. Last but not least, the StyleGAN model is easier to implement and faster to converge during optimization, compared to other latent vector based models such as Diffusion Models \cite{rombach2022high}. This enhancement module is referred to as the Vector-to-Unsupervised Mask-to-Image (V2UM2I) synthesis, and both UM2I and V2UM2I are UM-guided syntheses.

\subsubsection{Annotated CT Images from Unsupervised Mask}
We developed a pix2pix structure \cite{isola2017pix2pix} to synthesize annotated CT images from unsupervised masks. The annotated image generation is featured in a multi-task setting. Specifically, we hypothesize that the feature maps embedded in the hidden layers of the image synthesis model should contain rich information that can assist in accurate segmentation synthesis with only a limited number of annotated data. With this hypothesis, we developed a multi-task learning based training strategy to obtain synthetic images and segmentation masks simultaneously. The image synthesis task can help the convergence of segmentation tasks during the optimization and thus, reduce the training data required for segmentation. 

To further utilize the semantics of unlabeled data, we introduced a semi-supervised training scheme that could harness both labeled and unlabeled data. For images with ground truth annotations, the loss function for the generator was composed of two parts: (1) the image synthesis loss: the adversarial loss $L_{adv}^I$ computed by the image discriminator $D_2^I$ and the VGG loss $L_{vgg}$ and (2) segmentation synthesis loss: the adversarial loss $L_{adv}^S$ computed by the segmentation discriminator $D_2^S$ and the segmentation loss $L_S$. In our experiments, we used a combination of soft DICE loss and Cross Entropy loss. 

During each epoch of training, we first optimized the image discriminator and the segmentation discriminator, then optimized the annotated image generator. After training on a small, labeled dataset, we saved the parameters of $G_2$ and $D_2^S$. We then used these pre-trained models to generate pseudo-segmentations for the unlabeled images, and the confidence maps C for the pseudo-segmentations. During training, the model only trusted the segmented regions with high confidence and only computed the segmentation loss $L_S$ on these trusted regions. For images without ground truth annotations, the loss function was composed of the image synthesis loss and the masked segmentation loss. 
\begin{figure*}[t]
\centerline{\includegraphics[width=15cm]{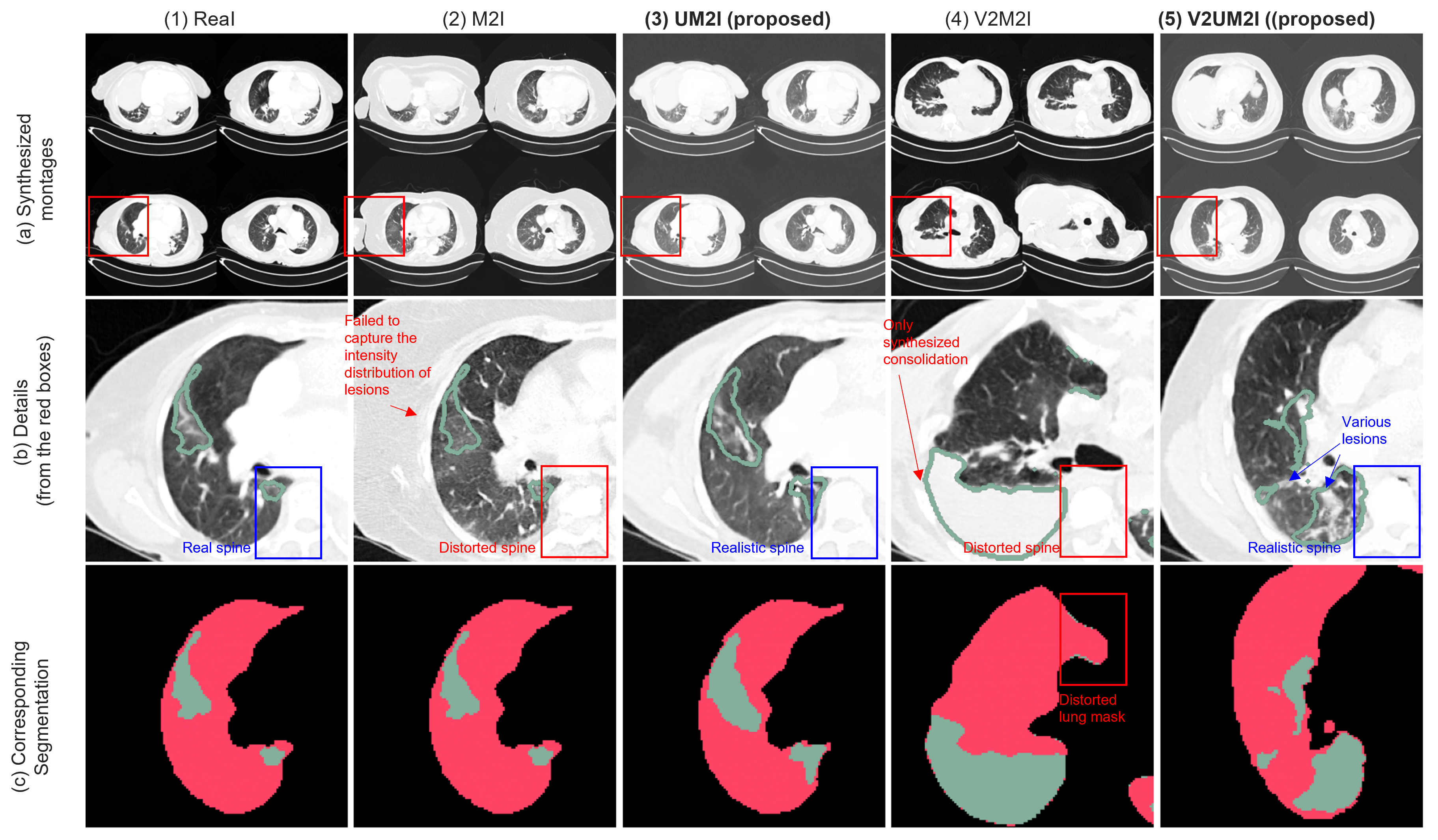}}
\caption{Showcases of synthetic images. Both M2I and UM2I results were obtained using real segmentation masks and real unsupervised masks, so we presented the ground truth segmentation masks and images in column 1. Our proposed algorithms UM2I (3) and V2UM2I (5) could produce segmentation masks simultaneously with the synthetic images. This figure illustrates that the segmentation mask is not a proper structural guidance for image synthesis, leading to unrealistic surrounding tissues and human bodies. The monotonous condition provided by segmentation masks could also lead to monotonous lesion pattern, as shown in (b4). In comparison, our algorithm could capture the opacity distribution of the lesions and could synthesize realistic COVID-related lesions.}
\label{fig:showcases}
\end{figure*}
\subsection{Stage 3: Model evaluation}
 In this study, we evaluated the quality of synthetic images in three aspects: fidelity, variety, and utility. In particular, we introduced MM-FID for the evaluation of fidelity and MM-STD for the evaluation of variety. 
 
\subsubsection{Fidelity: Multi-scale Multi-task FID (MM-FID)}
Fidelity is defined by the similarities between real and synthetic data distributions. The Fréchet Inception Distance (FID) \cite{heusel2017gans} was mostly used in many data synthesis tasks. The core idea of the FID score is to measure the similarities of distributions between features extracted from real and synthetic data. Because the FID score focuses on semantic features extracted from pre-trained models, it may mimic human perception of image fidelity. 

A common practice to compute FID score on images is: first, using an InceptionV3 \cite{szegedy2016rethinking} pre-trained on ImageNet \cite{deng2009imagenet} to extract feature maps from both real and synthetic images, and then model the distributions of these two feature map sets using two separate multi-dimensional Gaussian distributions. By observing the feature distributions, the image distributions can be approximated. Then the FID can be calculated by the squared Wasserstein metric between these two distributions. However, for one dataset, only a single observation of feature distribution can be made, thus the computed FID score may be biased according to the feature extractor model. In addition, the CT images are grey-scale images, and a model that is pre-trained on RGB images may fail to extract representative features of CT images. Thus, to provide a robust analysis of the synthesis fidelity, in this paper, we measured the FID score under different resolutions with different feature extraction methods. 

More specifically, we resized the images into four different resolutions, ranging from $2^7\times2^7$ to $2^{10}\times2^{10}$. Then we used 4 different models pre-trained on different tasks to extract features. \textcolor{black}{We carefully selected the lowest scale for our multi-scale FID as $2^7$ because we discovered that images under the resolution of $2^7$ failed to provide detailed information for human observers (particularly for our clinician collaborators) to discriminate whether the CT images were real or synthetic. Then we expanded the multi-scale grid in the range of 7 to 10 in the power of 2.} The tasks include ImageNet classification \cite{deng2009imagenet}, which is a common practice for FID score, sex, and age classifications on our datasets, and a discriminatory task that classified real images from synthetic images. The latter three tasks are more relevant to our dataset and thus could produce highly representative features for fidelity evaluation.

By doing so, we could obtain 16 sets of features from the images, and each set of features had an FID score with the real features. Then, we could interpret the FID scores statistically and could provide a fidelity measurement of synthetic images with statistical significance. Since FID calculated on different feature distributions might vary dramatically, we normalized the final FID score into $[0,1]$. 

\subsubsection{Variety: Multi-scale Multi-task STD (MM-STD)} 
The variety of datasets can be measured in two aspects: intra-class variety and inner-class variety \cite{deng2009imagenet}. The intra-class variety requires a classification label for each image in the dataset and measures the number of images for each classification group, e.g., the male to female ratio can indicate an intra-class variety in medical datasets. The inner-class variety measures the variety of images that belong to the same classification group. In the ImageNet dataset \cite{deng2009imagenet}, the inner class variety is defined by the lossless JPEG file size of the average image for each classification group. The authors justified this definition by presuming that a dataset containing diverse images will result in a blurrier average image, thus will reduce the lossless JPEG file size. 

 However, it is difficult to assign classification labels for synthetic images once the synthesis model is converged. Due to this reason, the variety assessment for synthetic images is scarce in the literature \cite{borji2019pros}. To mimic the human perception of the image variety \cite{heusel2017gans}, in this paper, we proposed to measure the variety by the standard deviations (STDs) of features extracted with pretrained networks. As aforementioned, we could obtain 16 sets of features for one dataset and then compute 16 STDs for this dataset. By doing so, we could statistically compare the varieties between real and synthetic images using MM-STD. In addition, we adopted the concept of JPEG file size of the average image in ImageNet \cite{deng2009imagenet} for the variance visualization. 
\subsubsection{Utility: Data Augmentation Performance}
We defined the utility score as the improvement of segmentation models after adding synthetic images to the training dataset. We tested the data augmentation performance of synthetic images under three circumstances, where the number of available labeled data varied: 
\begin{enumerate}
    \item 	Insufficient data: only 300 CT scans from 65 patients with labeling are provided. 
	\item A moderate amount of data: only 600 CT scans from 160 patients with labeling are provided. 
	\item Sufficient data: 1500 CT scans from 728 patients with labeling are provided. 
\end{enumerate}

\begin{figure}[h]
\centerline{\includegraphics[width=\columnwidth]{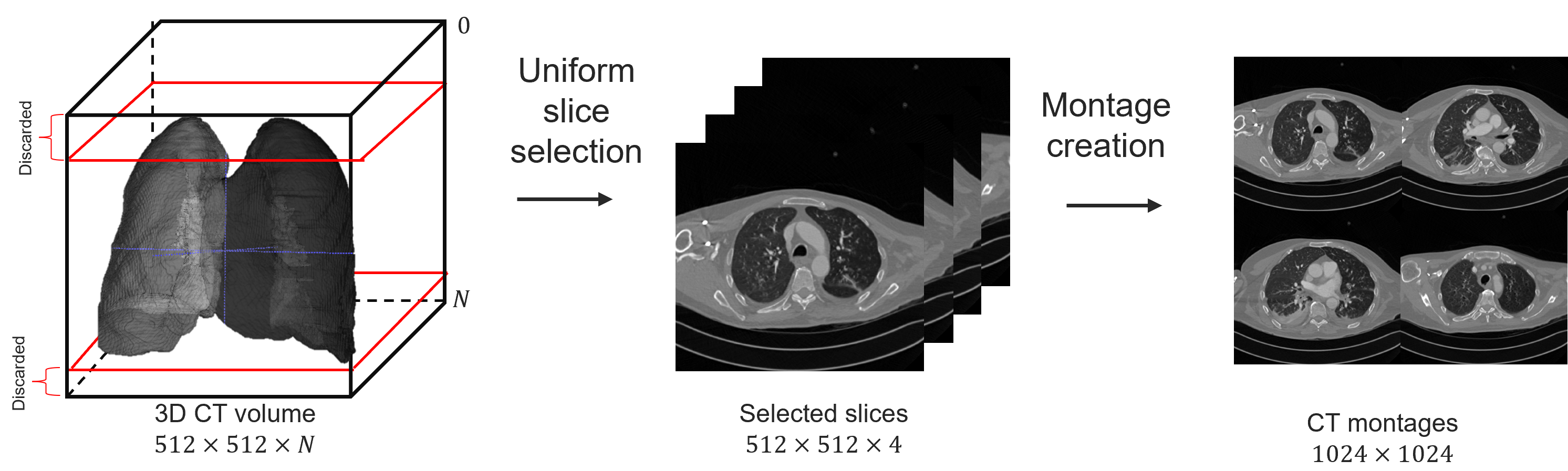}}
\caption{The generation process of CT montages. The top and bottom slices that had lung regions $<10\%$ were removed.}
\label{fig:preprocess}
\end{figure}

Under each circumstance, we generated 1000 synthetic montages (see Fig. \ref{fig:preprocess}) from the available data. We used the montage representation to 1) balance the dilemma between the GPU costs for 3D CT volumes and the information consistency among slices \cite{walsh2018deep}, and 2) handle the slice number variances in our dataset. We selected 4 2D CT slices and tiled them in a $2\times2$ manner. We then trained one segmentation UNet with the available data only and one segmentation UNet with both available real data and 1000 synthetic data. Both models were inferenced on an independent test dataset, and the improvement of segmentation performance was defined as the utility score.

\section{Experiments}
\label{sec:experiments}
\subsection{Data and Experimental Settings}
The data we used was obtained from the National COVID-19 Chest Image Database (NCCID) \footnote{\url{https://nhsx.github.io/covid-chest-imaging-database/}}. The datasets were acquired from 11 UK hospitals that covered a large variation in the patient population and data quality. Most patients (83.48\%) had one scan, and the remaining had $2-9$ scans. The raw data were reconstructed using $1-4$ different reconstruction kernels, which means that the image quality and appearance varied and were not harmonized across imaging sites. This imposed additional challenges in our synthesis algorithm. Overall, 3292 CT scans from 1072 patients were included in this project. We used an automatic segmentation tool (Contextflow \cite{hofmanninger2020automatic}) to first segment the lung and COVID-related lesions and manually adjusted the segmentation results for each volume under the supervision of our clinical radiologists. COVID-related lesions include both GGO and consolidation. We applied a Hounsfield Unit (HU) window ranging from -1500 to 100 for all CT volumes. The pixel spacing of these images was $[0.6875 mm, 0.6875 mm]$, leading to an image array size of $512\times512$ for each 2D slice in these 3D volumes. The total number of slices in each volume varied in the range of $[36, 424]$. 

Since the whole volumetric synthesis consumes large GPU memories, which could reduce the efficiency of the proposed method, we developed a CT montage based framework, which is composed of 4 axial high-resolution CT image slices, as the image representation for our synthetic data. The CT montage presentation has been used in clinical lung fibrosis analysis \cite{walsh2018deep} and has been proven to maintain an effective balance between the computation efficiency and the cross-slice information. In addition, it sufficiently addresses the varied slice thickness problems and avoids redundant inter-slice interpolation. 

For each scan, the top and bottom slices that had lung regions $<10\%$ were discarded because of limited tissue information. We then divided the remaining slices into 4 equal clusters according to their axial position. For each cluster, one slice was randomly selected, and we tiled the 4 selected slices from all clusters into a 4-image 2D CT montage with a resolution of $1024\times 1024$. For each patient, one CT montage was generated. During our experiments, we randomly selected 1500 CT montages for training, 1000 CT montages for validation, and 785 CT montages for independent testing. 

\subsection{Experimental Settings}  
\textbf{Network Parameters.} Regarding the superpixels, we used $M=512$, $t=50$ for all our comparison models because they could produce the best synthetic CT montages according to our ablation experiments in Section V. B. All our models were implemented with Python and PyTorch (version=1.9.0) and trained on 4 RTX 3090 CUDA devices. For $G_1$, we developed a StyleGAN based architecture to generate the unsupervised masks for images. We used the progressive training strategy and truncation strategy of the styleGAN. The overall depth of the styleGAN used was 7. The epoch and batch size for each depth were [8, 16, 16, 32, 32, 64, 64] and [128, 128, 128, 64, 32, 16, 8]. The learning rate was 0.003.

The architecture for $G_2$ was composed of 4 down-sampling, 9 residual, and 4 up-sampling blocks. Each block contains one convolutional layer, one batch normalization layer, and one activation layer. The final loss function for $G_2$ was composed of the cross-entropy loss $L_{ce}$, the inception loss $L_{vgg}$ computed from a pre-trained VGG network \cite{johnson2016perceptual} and the discriminator loss $L_{adv}$ from the discriminator. M2I models used the same architecture as UM2I models, while the input and output for M2I models were the segmentation masks and the synthetic images, and $L_{vgg}$ and $L_{adv}$ were used to optimize the M2I models. During training, the segmentation masks and UMs were not paired with the real images. 

For the comparison of the segmentation performance, we used an in-house UNet model with 4 downsampling blocks. We trained our model in parallel on 4 CUDA devices with a batch size of 16 and a learning rate of 0.01. The loss function for our segmentation UNet was a combination of soft DICE loss and cross-entropy loss. 

\textcolor{black}{The peak GPU memory consumption observed during training was 8.3 GB for 8 images per GPU. In the second stage, the peak GPU memory consumption observed during training was 22.8 GB for 4 images per GPU. }

 \textbf{Pre-trained Models.} In this paper, we proposed to use multiple feature distributions to measure the fidelity and variety of synthetic images statistically. For each dataset, 16 sets of features were extracted from different resolutions with different pretrained models. Overall, 4 resolutions ranged from $2^7$ to $2^{10}$ were used, and 4 InceptionV3 \cite{szegedy2016rethinking} models were used to extract the image features. The first InceptionV3 model $P_1$ was pre-trained on the ImageNet dataset and has been widely used in FID score calculations. The second and third InceptionV3 model $P_2$ and $P_3$ were pre-trained on our training dataset, classifying gender and age. \textcolor{black}{We categorized the age into four groups according to different responses of these age groups to COVID. The four groups are (0, 44], (44, 64], (64, 74], (74,100].} The last Inception V3 was pre-trained on both real images and synthetic images. For both V2M2I and V2UM2I, we generated 1000 synthetic images. The optimization target is to discriminate whether the images are real, synthesized from V2M2I, or synthesized from V2UM2I. \textcolor{black}{The latter three models ($P_2$, $P_3$, $P_4$) were trained under the resolution of $2^10×2^10$}.  During the training of the latter 3 models , data augmentation techniques including cropping, shifting, rotation, and noise-adding were used to extract robust feature representations.

\subsection{Experimental Results}

Examples of synthetic images and lesion segmentations can be found in Fig. \ref{fig:showcases}. As is shown in columns (1-3), our algorithm could capture the CT signal intensity distributions of lesions, as well as adequate constraints on all human anatomies, including those non-ROI regions. In this section, we demonstrated the efficiency of the proposed UM2I and V2UM2I for image synthesis. We majorly compared our algorithm (UM2I, V2UM2I) with segmentation mask-guided synthesis algorithms (M2I, V2M2I) on three aspects: fidelity, variety, and utility. 
\begin{figure}[t]
\centerline{\includegraphics[width=9cm]{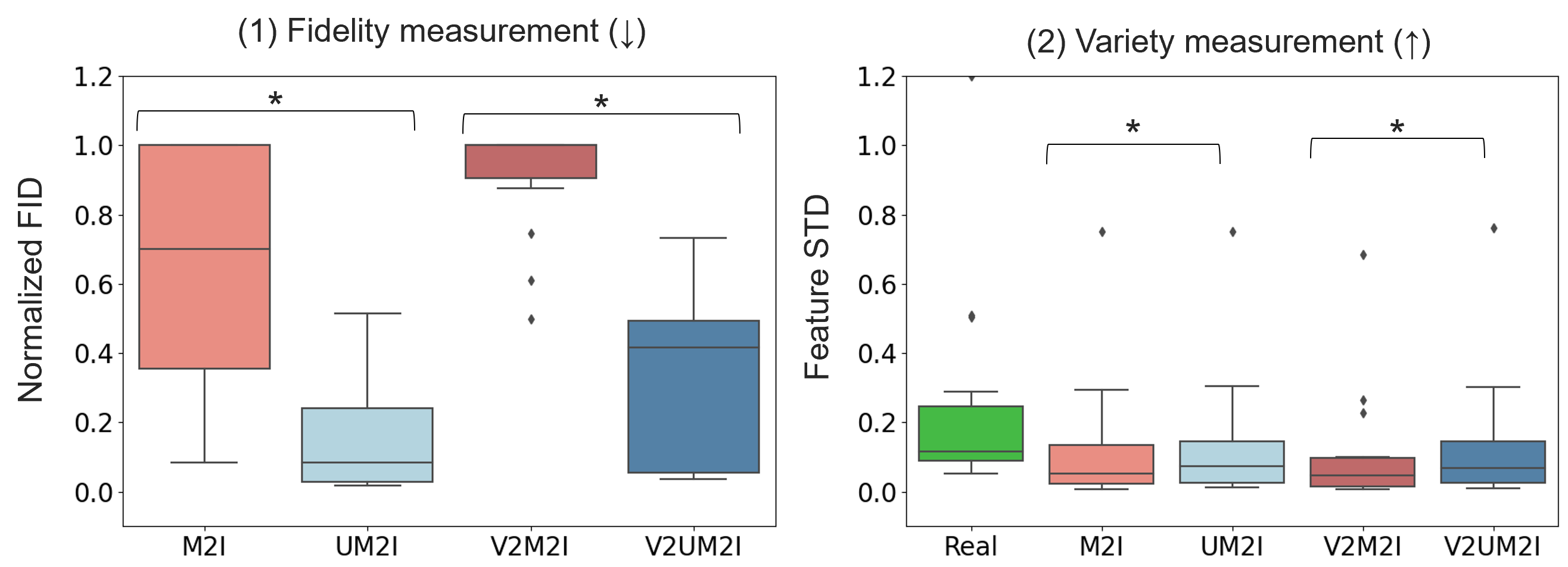}}
\caption{Box plots of fidelity and variety from different sets of synthetic images. The fidelity is measured by FID scores between feature distributions, and the variety is defined by the STD of extracted features. Here, * indicates a significant difference ($p<0.05$) between the two groups.}
\label{fig:boxplots}
\end{figure}

\subsubsection{Fidelity}
The mean and variance of FID scores of our proposed UM-guided methods (UM2I, V2UM2I) and the comparison segmentation-guided methods (UM2I, V2UM2I) are shown in Fig. \ref{fig:boxplots} (1). Our UM2I method demonstrated a higher similarity with real datasets. \textcolor{black}{It should be noted that in our work, we used FID to evaluate the fidelity, but the proposed multi-task multi-scale approach can also be applied to other popular evaluation metrics such as Inception Score \cite{salimans2016improved} and KID \cite{binkowski2018demystifying}.}

\subsubsection{Variety}
We quantified the variety of our dataset as the standard deviation of images extracted from pretrained models. As shown in Fig. \ref{fig:boxplots} (2), our UM-guided synthesis can generate images that are significantly more varied than segmentation guided synthesis. 

\subsubsection{Utility}
We evaluated the utility of synthetic images indirectly by measuring the performance improvement when adding synthetic images to the training datasets. We tested the data augmentation performance of synthetic images under three circumstances including insufficient data (300 montages), a moderate amount of data (600 montages), and sufficient data (1500 montages). 
\begin{table}[h]
    \caption{The segmentation DICE score for lung and lesions \textcolor{black}{trained on real data and with different numbers of synthetic data}. Here, * indicates $p<0.05$ for Wilcoxon Signed Rank Test and the statistically significant difference compared with the model trained \textcolor{black}{only} on real data, respectively.}
    \centering
\begin{tabular}{lllll}
\hline
Region & Method & Insufficient & Moderate & Sufficient \\\hline
\multirow{3}{*}{Lung} & Real & 0.91±0.13 & 0.91±0.12 & 0.93±0.09 \\\hline
 & +V2M2I & 0.86±0.20$^*$    & 0.92±0.11$^*$ & 0.93±0.09 \\\hline
 & +V2UM2I & 0.92±0.10$^*$ & 0.93±0.10$^*$ & 0.94±0.09$^*$ \\\hline
\multirow{3}{*}{Lesion} & Real & 0.45±0.23 & 0.58±0.25 & 0.63±0.26 \\\hline
 & +V2M2I & 0.41±0.25$^*$ & 0.57±0.25 & 0.63±0.32$^*$ \\\hline
 & +V2UM2I & 0.51±0.25$^*$ & 0.62±0.26$^*$ & 0.64±0.26$^*$\\\hline
\end{tabular}

\label{tab:augmentation}
\end{table}

To ensure a fair comparison, the V2UM2I was trained in a fully supervised manner. We generated 1000 synthetic montages from V2M2I and V2UM2I models with a different number of real images. \textcolor{black}{We then trained three segmentation UNets on 1) real data and 2) real data with 1000 synthetic montages from the V2M2I model and 3) real data with 1000 synthetic montages from the V2UM2I model for data augmentation. A Wilcoxon Signed Rank Test was performed to validate the significance of difference. }As shown in Table \ref{tab:augmentation}, the UM-guided synthesis can better boost the performance of the downstream segmentation with 300 montages when compared to conventional conditional synthesis models. 

\subsubsection{Efficiency of multi-task training}
\begin{table}[h]
\caption{Comparison of the performance of independently synthesized images and segmentation masks versus multi-task synthesized images and segmentation masks.}
\centering
\begin{tabular}{llll}
\hline
Method & Fidelity (↓) & Lung DICE  (↑) & Lesion DICE (↑) \\ \hline
Image only & 0.57 ± 0.34 & / & / \\ \hline
Segmentation only & / & 0.84 ± 0.17 & 0.51 ± 0.19 \\ \hline
Proposed & 0.43 ± 0.38 & 0.92 ± 0.10 & 0.5 1± 0.25 \\ \hline
\end{tabular}
\label{tab:indepent}
\end{table}
\textcolor{black}{In our work, we found that the image synthesis and image segmentation tasks have similarities in terms of their need for abundant semantic information to produce satisfactory results. Based on this observation, we incorporated both tasks into a multi-task setup. To validate this approach, we conducted additional experiments using the V2UM2I pipeline to synthesize both images and segmentation masks independently and evaluated their performance using the fidelity score for image synthesis and the DICE score for segmentation augmentation (as shown in Table \ref{tab:indepent}). }

\subsubsection{Efficiency of semi-supervised training}
\textcolor{black}{To assess the efficacy of our proposed semi-supervised learning approach, we conducted experiments to compare the segmentation augmentation performance using V2UM2I synthesized data, as in Table \ref{tab:semi}. For example, in the data insufficiency scenario, we trained the UM2I component in both a fully supervised manner using only 300 labeled images and a semi-supervised manner using 300 labeled images and 1200 unlabeled images. We then generated 1000 synthetic images and compared the segmentation augmentation performance by training the network with real data, which included 300 labeled real images and 1000 labeled synthetic images. This approach can be extended to the moderate scenario, where 600 labeled images were available.}

\begin{table}[h]
\caption{Comparison of segmentation augmentation performance between fully supervised and semi-supervised training.}
\begin{tabular}{lllll}
\hline
 Region & Method & Insufficient & Moderate & Sufficient \\ \hline
Lung & Fully-supervised & 0.89 ± 0.13 & 0.91 ± 0.11 & 0.94 ± 0.09 \\ \hline
 & Semi-supervised & 0.92 ± 0.10 & 0.93 ± 0.10 & / \\ \hline
GGO & Fully-supervised & 0.47 ± 0.10 & 0.58 ± 0.20 & 0.64 ± 0.26 \\ \hline
 & Semi-supervised & 0.51 ± 0.25 & 0.62 ± 0.26 & / \\ \hline
\end{tabular}
\label{tab:semi}
\end{table}

\section{Discussions}

\subsection{UM-Guided Synthesis Can Reduce the Mode Collapse Problem}

According to our experiments, the loss of variety is inevitable during image synthesis. In Fig. \ref{fig:boxplots} (2), the varieties of synthetic images from all conditional synthesis algorithms are smaller than real images. However, we claimed that introducing intensity guidance can address and improve the loss of variety in synthetic images. To ensure a fair comparison, we compared images synthesized from V2M2I models and V2UM2I models.
\subsubsection{Variety of Image Layouts}

\begin{figure}[t]
\centerline{\includegraphics[width=9cm]{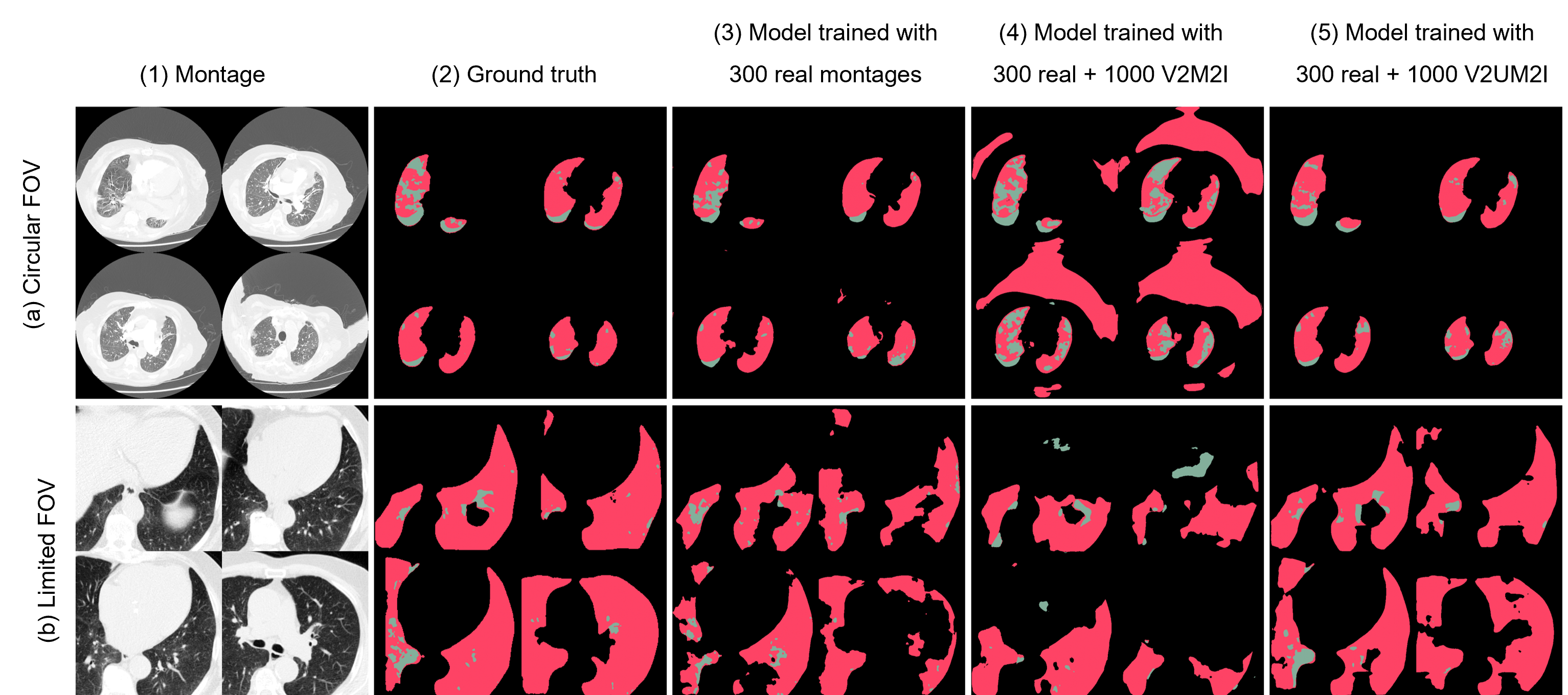}}
\caption{In data insufficient scenarios, the segmentation performance of models trained on 300 real montages (3), with 1000 V2M2I montages or with 1000 V2UM2I montages. As is presented in subfigures (4a) and (4b), the V2M2I failed to synthesize images with circular FOV and images with limited FOV, thus leading to a worse segmentation performance on these images in the limited data scenario (300 montages). }
\label{fig:segmentation}
\end{figure}

To visualize the data distributions, we used features derived from the pretrained model $P_4$ under the resolution of $2^{10}\times2^{10}$ because the target of $P_4$ is to discriminate different image types, and t-distributed Stochastic Neighbor Embedding (T-SNE) \cite{hinton2002stochastic} was used to reduce the feature dimensions. 
 \begin{figure}[h]
\centerline{\includegraphics[width=9cm]{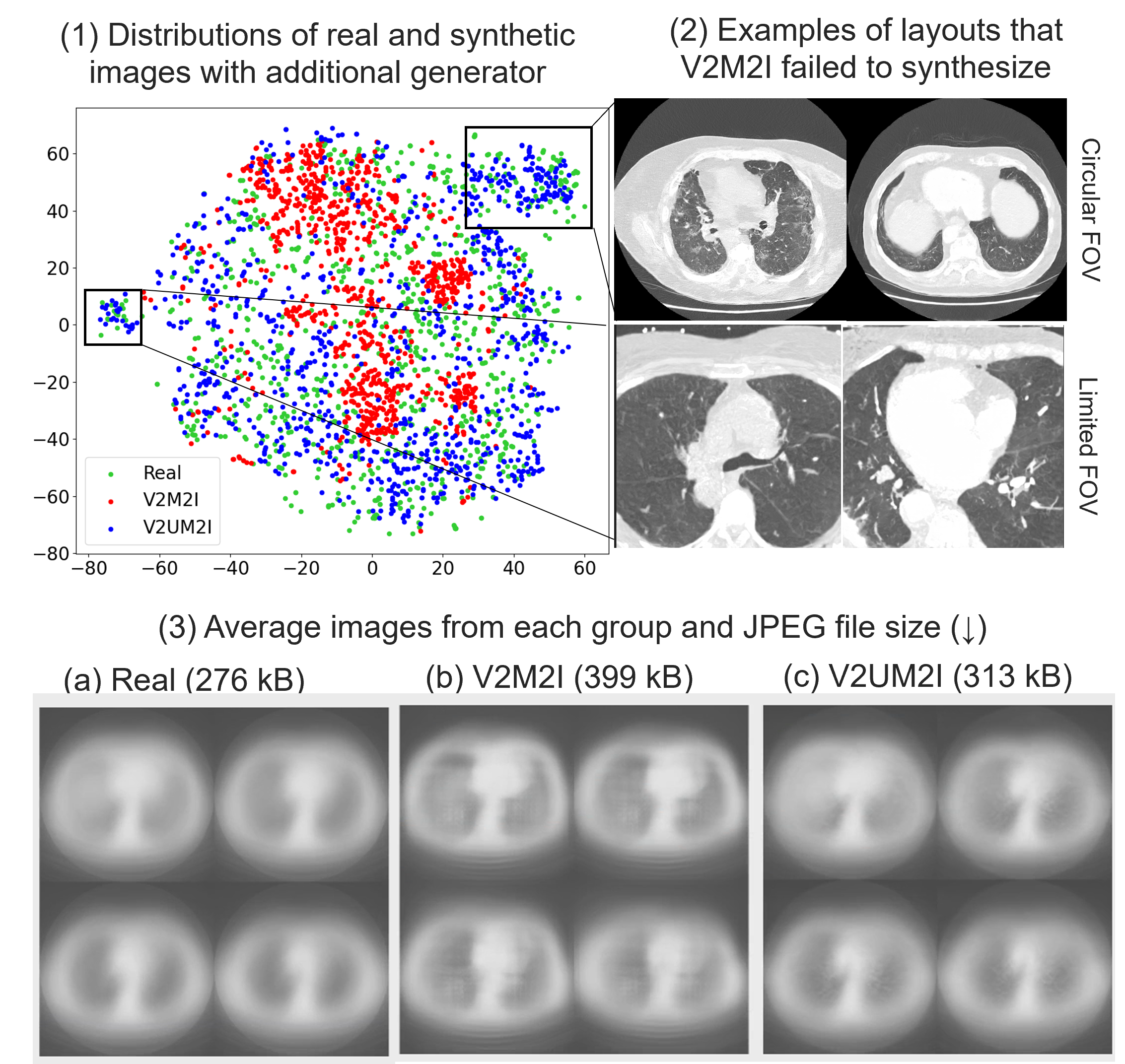}}
\caption{Visualization of data distributions in (1), examples of two clusters where the V2M2I model failed to synthesize in (2), and the average image for each group in (3). The JPEG file size was calculated to demonstrate the variety of each group, and a lower value indicates a higher variety.  }
\label{fig:layout}
\end{figure}
We discovered two clusters in the image space that the V2M2I model failed to cover, as shown in the black boxes in Fig. \ref{fig:layout} (1). One cluster is located between $x=[-80,-70]$ and $y= [0, 20]$, and the other is located between $x=[20, 40]$ and $y=[-80,-60]$. We randomly selected two V2UM2I synthesized images from each cluster (Fig.  \ref{fig:layout} (2)). This indicates that the V2M2I model failed to synthesize CT montages with different FOVs, but our V2UM2I was able to perform these syntheses. Our real dataset contains images reconstructed with different kernels, thus producing images with limited FOVs and circular FOVs. As a result of this mode collapse, the V2M2I model failed to improve the lesion segmentation accuracy on images with limited FOVs in data insufficient scenarios (Fig. \ref{fig:segmentation}). In addition, we visualized the average images of different groups (Fig. \ref{fig:layout} (3)). By calculating the lossless of JPEG file sizes of the average images, we further proved that our V2UM2I algorithm could preserve the data variety (313 kB) compared to segmentation-guided synthesis (399 kB). 
\begin{table}[h]
\caption{Comparison of Average JPEG File Size and noise levels of real and synthetic datasets using NLM and BM3D algorithms.}
\begin{tabular}{lllll}
\hline
Synthetic method & \begin{tabular}[c]{@{}l@{}}File size of\\ average image (↓)\end{tabular} & NLM (↓) & BM3D (↓) \\ \hline
Real & 276 kB & 0.02 ± 0.03 & 0.13 ± 0.14 \\ \hline
UM2I & 399 kB & 0.21 ± 0.16 & 0.11 ± 0.10 \\ \hline
V2UM2I & 313 kB & 0.15 ± 0.15 & 0.12 ± 0.07 \\ \hline
\end{tabular}
\label{tab:noise}
\end{table}

\textcolor{black}{The JPEG file size of average image can be influenced by both the diversity of the images and the noise level. To determine the impact of these factors on the file size, we evaluated the noise level of synthetic images using two commonly used methods, NLM \cite{buades2005non} and BM3D \cite{dabov2007image}. The results, shown in Table \ref{tab:noise}, provide further evidence for the effectiveness of our proposed algorithm. Under the same level of noises, the average image of images synthesized from our proposed algorithm has the lowest JPEG file size, indicating a highest variety.}

\subsubsection{Variety of Synthetic Lesions }
 \begin{figure}[t]
\centerline{\includegraphics[width=9cm]{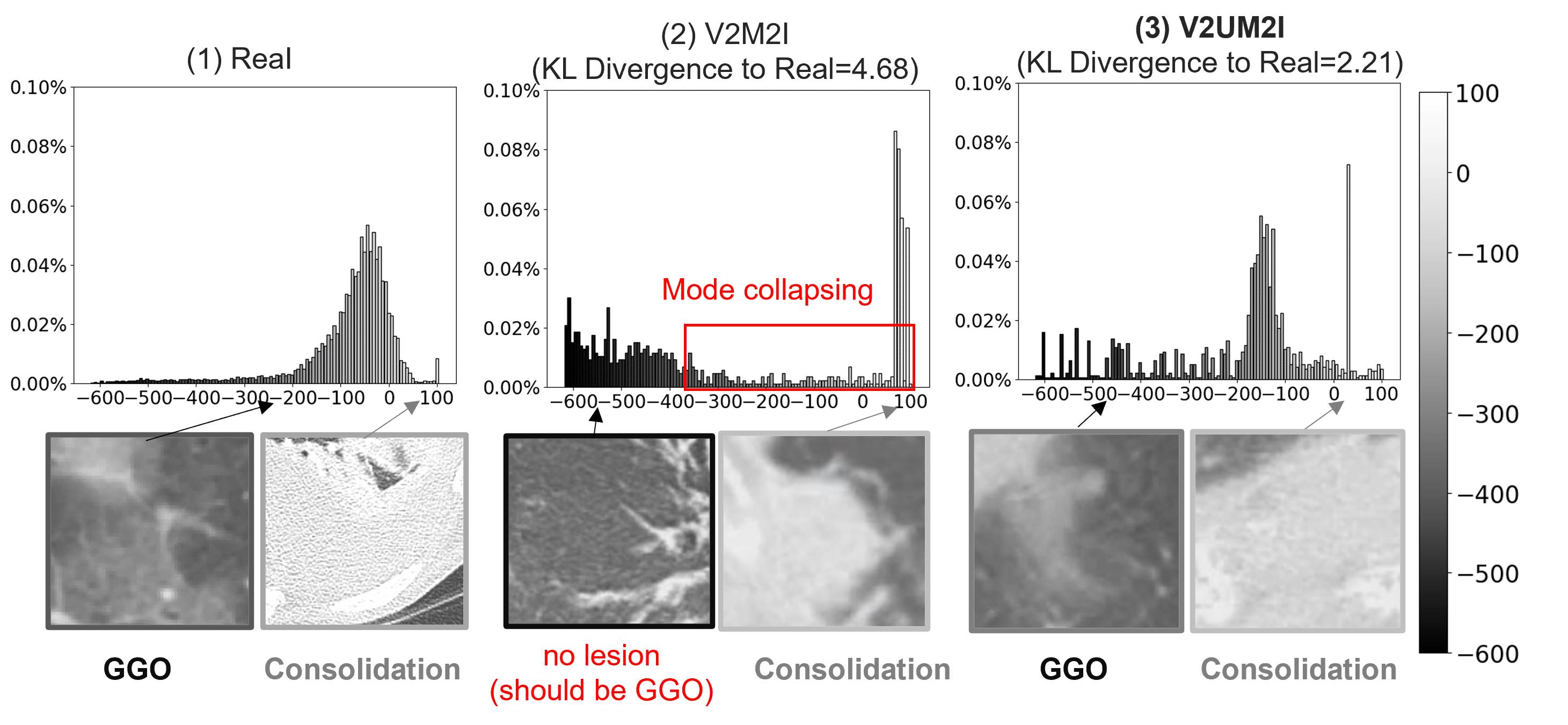}}
\caption{Histograms and examples of synthetic lesions from each group. The V2M2I method failed to synthesize lesions with HU values ranging from -300 to 0 (the GGO pattern). }
\label{fig:lesion}
\end{figure}

We also discovered a mode collapse problem in synthetic lesions from segmentation guided synthesis. We plotted the histograms of HU values of the synthetic lesions in Fig. \ref{fig:lesion}. The segmentation-guided models have a severe mode collapse problem, in which only lesions with high HU values (e.g., consolidation patterns) were synthesized. We computed the Kullback-Leibler divergence (KL divergence) of the HU value distributions, and our intensity-guided networks could model the original lesion distributions faithfully with a lower KL divergence (2.21 compared to 4.68 of the V2M2I model). 

\subsection{Parameter selection for superpixel generation}
\subsubsection{Parameter selection for SLIC-based superpixel}
Two hyperparameters are influencing the generation of unsupervised masks: the number of superpixels $M$ and the intensity threshold $t\in (0,255]$. Because we clustered the superpixels based on their intensity values, the intensity threshold $t$ plays a more crucial role during the unsupervised mask generation. Thus, in this section, we investigate the effects of the threshold value $t$ on the quality of synthesized images. 
 \begin{figure}[t]
\centerline{\includegraphics[width=9cm]{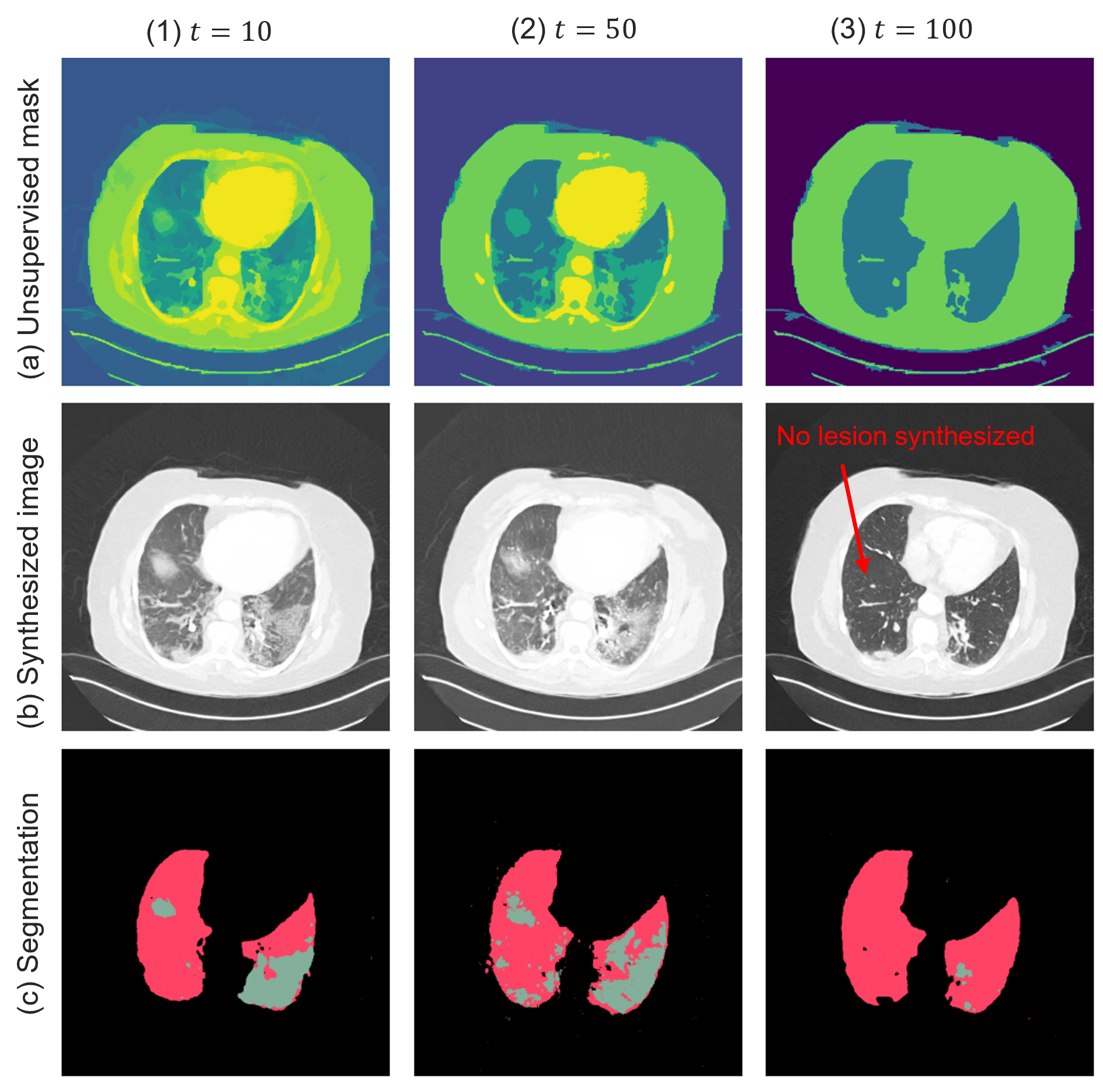}}
\caption{Unsupervised masks with different $t$ values (a) and their corresponding synthetic images (b) and annotations (c). As shown from (a1-3), a higher $t$ value indicates fewer details provided by the unsupervised masks. Higher $t$ values, although simplifying the image synthesis task, might synthesize images that were not the same as the corresponding real images (b3). In this case, during training, the segmentation masks from real images might fail to provide adequate supervision for segmentation synthesis. }
\label{fig:tr}
\end{figure}

All unsupervised masks were normalized into 0-255 at first. As is shown in Fig. \ref{fig:tr} (a), a smaller value ($t \leq 50$) resulted in more super-clusters and provided detailed structure guidance. Detailed guidance with lower $t$ values preserved the structure of corresponding real CT montages and allowed the synthetic images to bear more resemblance to their corresponding input real CT montages. By doing so, we could assure that the semi-supervision was accurate during training. In comparison, higher $t$ values (Fig. \ref{fig:tr} (b3) ) might synthesize images that were not the same as the corresponding real images, reducing the accuracy of supervision during the segmentation synthesis. 
However, because the unsupervised masks with lower $t$ values contained too many details, synthesizing realistic unsupervised masks with more details was more difficult for the V2UM model (Table \ref{tab:tr}). In extreme conditions where the input unsupervised masks are as detailed as real images, our UM-guided synthesis algorithm becomes a StyleGAN for image synthesis and a self-auto encoder for image segmentation. The direct synthesis of high-resolution images led to mode collapse problems and unrealistic synthetic images. In Table \ref{tab:tr}, we used different background colors to demonstrate the performance: a red cell indicates an unacceptable poorer performance, and a green cell indicates an acceptable performance. According to Table \ref{tab:tr}, we used $t=50$ for all our experiments to address this dilemma between the quality of synthetic segmentations and the quality of synthetic images.
\begin{table}[h]
\caption{The performance of synthetic images with different input structural guidance.  Here $t=0$ indicates an extreme condition where the V2UM branch was synthesizing images directly. Red cells present results that are significantly poor, and green cells present comparable performance.}
\centering
\begin{tabular}{llll}
\hline
$t$ & Fidelity   score (↓) & Variety   score (↑) & Leision DICE (↑) \\\hline
0 & \cellcolor[HTML]{FF0000}0.85 ± 0.26 & \cellcolor[HTML]{FF0000}0.11 ± 0.17 & \cellcolor[HTML]{00B050}0.63 ± 0.27 \\\hline
10 & \cellcolor[HTML]{FF0000}0.75 ± 0.29 & \cellcolor[HTML]{00B050}0.13 ± 0.19 & \cellcolor[HTML]{00B050}0.65 ± 0.20 \\\hline
50 & \cellcolor[HTML]{00B050}0.43 ± 0.38 & \cellcolor[HTML]{00B050}0.14 ± 0.18 & \cellcolor[HTML]{00B050}0.64 ± 0.26 \\\hline
100 & \cellcolor[HTML]{00B050}0.33 ± 0.36 & \cellcolor[HTML]{FF0000}0.11 ± 0.18 & \cellcolor[HTML]{FF0000}0.51 ± 0.31\\\hline
\end{tabular}
\label{tab:tr}
\end{table}

\subsubsection{Comparison of superpixel generation algorithm}
\textcolor{black}{In our study, we utilized the SLIC algorithm to generate superpixels. In this section, we compared its performance to other state-of-the-art superpixel generation methods (Table \ref{tab:sp}), including Felzenszwalb and Huttenloch's graph-based algorithm (FSZ) \cite{felzenszwalb2004efficient} and SP-FCN \cite{yang2020superpixel}, which uses back-propagation optimization. We evaluated the generated superpixel masks using the under-segmentation error metric \cite{levinshtein2009turbopixels} and measured the synthetic image fidelity score. Then, we employed the V2UM2I synthesis pipeline and trained three segmentation networks using both real and synthesized images generated from different superpixel masks. The performance of the augmented segmentation masks was evaluated using the DICE score. Although SLIC had a higher under-segmentation error, we did not observe a significant difference in the performance of the synthesized segmentation masks in downstream tasks.}

\begin{table}[h]
\caption{The performance of synthetic images unsupervised masks generated from different algorithms. Only 300}
\centering
\begin{tabular}{p{1cm}lllll}
\hline
\begin{tabular}[c]{@{}l@{}}Superpixel \\ method\end{tabular} & \begin{tabular}[c]{@{}l@{}}Under-segment \\ error $\times 10^{-2}$ (↓)\end{tabular} & \begin{tabular}[c]{@{}l@{}}Fidelity\\score (↓)\end{tabular} & \begin{tabular}[c]{@{}l@{}}Lung \\ DICE  (↑)\end{tabular} & \begin{tabular}[c]{@{}l@{}}Lesion \\ DICE (↑)\end{tabular} \\ \hline
SLIC & 10.70 ± 5.09 & 0.43 ± 0.38 & 0.94 ± 0.09 & 0.64 ± 0.26 \\ \hline
FSZ & 8.70 ± 4.79 & 0.44 ± 0.27 & 0.93 ± 0.09 & 0.61 ± 0.30 \\ \hline
SP-FCN & 8.56 ± 5.22 & 0.81 ± 0.31 & 0.91 ± 0.10 & 0.64 ± 0.27 \\ \hline
\end{tabular}
\label{tab:sp}
\end{table}

\subsection{UM-Guided Synthesis for Image Lesion Editing }
Here, we also demonstrated the potential applications of our UM-guided synthesis in image lesion editing in both simple lesion patterns such as COVID-related lesions and more complex lesion patterns such as the usual interstitial pneumonia (UIP) pattern. 

 \begin{figure}[h]
\centerline{\includegraphics[width=9cm]{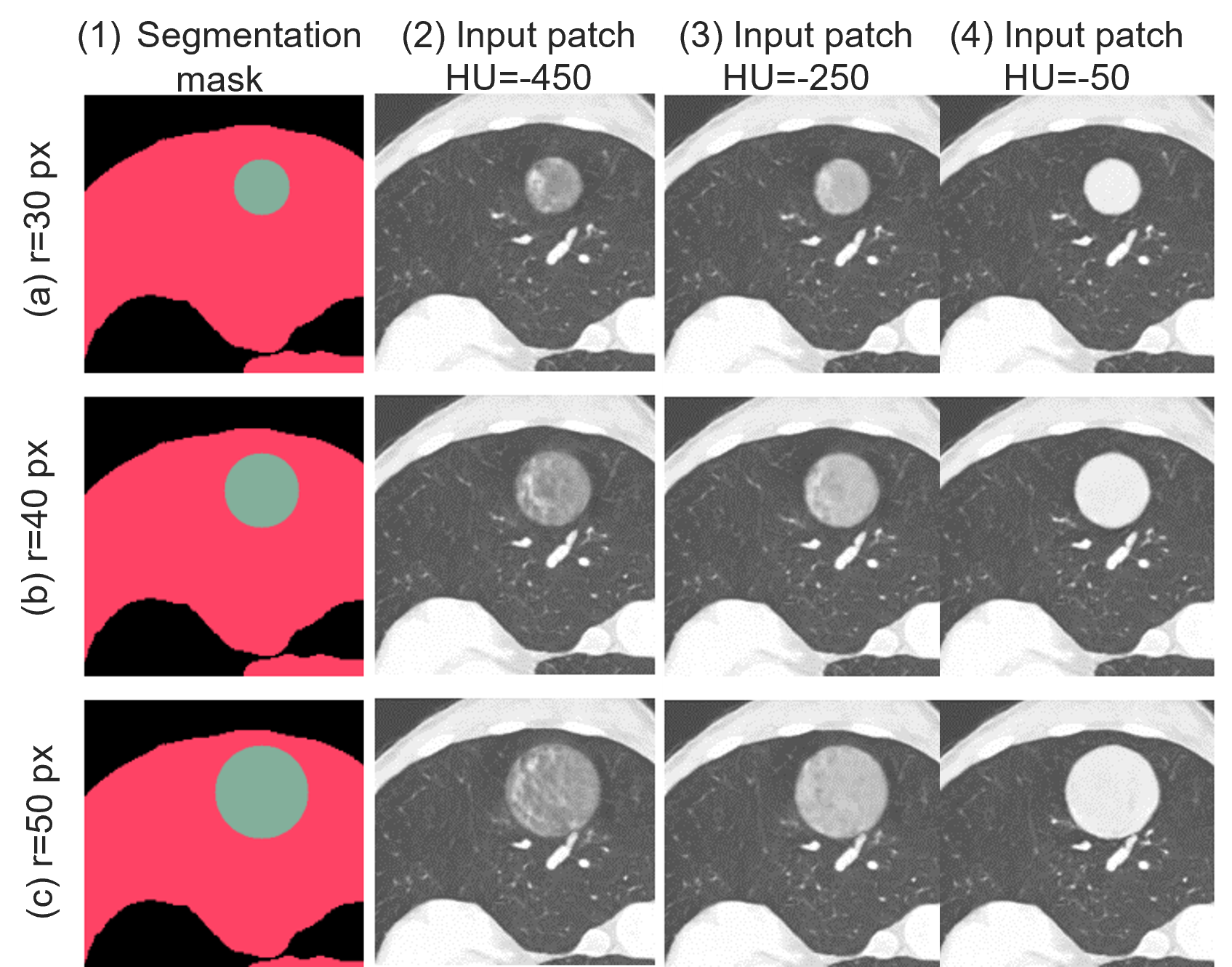}}
\caption{Synthetic lesions with different shapes from the UM2I model and the segmentation masks. We can use the intensity value to edit the appearance of lesion patches in UM2I algorithms to generate different types of lesions in the lung, from GGO (2) to consolidation (4).}
\label{fig:editing}
\end{figure}
 \begin{figure}[h]
\centerline{\includegraphics[width=9cm]{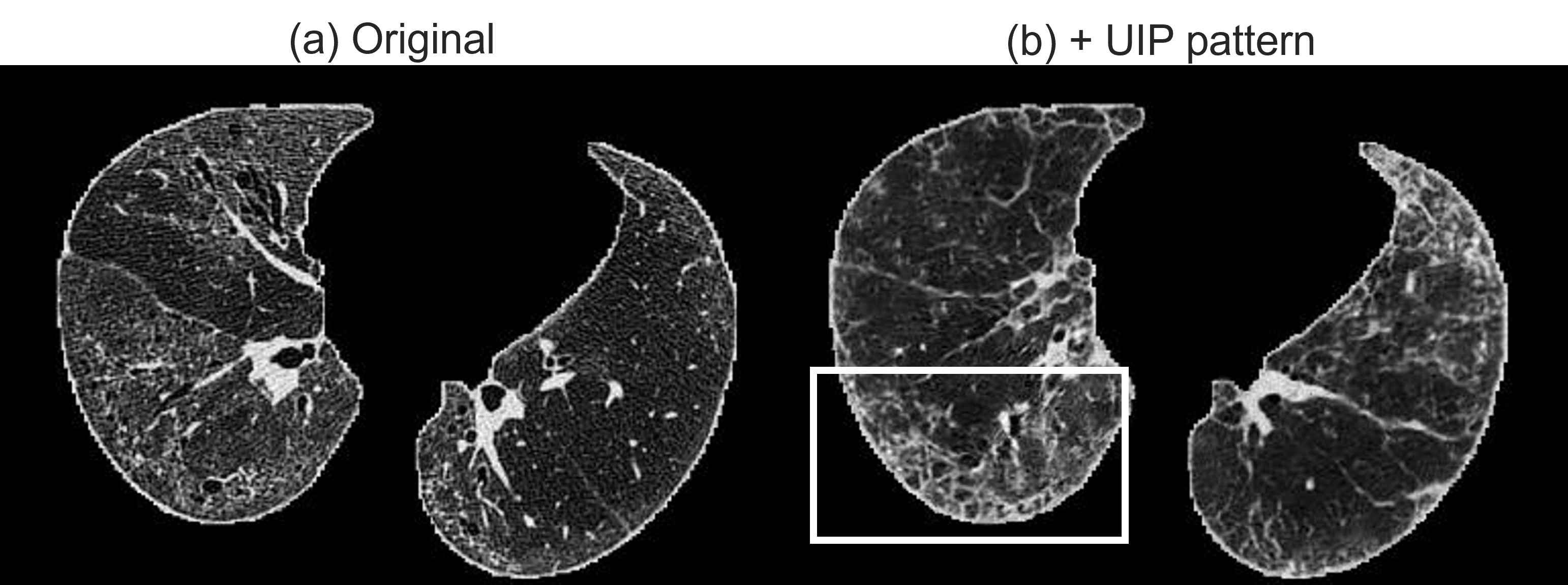}}
\caption{Synthetic UIP cases (b) from non-UIP cases (a) (the alternative diagnosis in the 2018 ATS/ERS/JRS/ALAT clinical practice guideline \cite{raghu2018diagnosis}).}
\label{fig:uip}
\end{figure}
\textbf{COVID-Related Lesion Editing.}The unsupervised masks also allow for image editing that will facilitate disease progressive modeling. Nearly 60\% GGOs will grow or increase their solid components \cite{kobayashi2013management}, and by interactively modeling the lesions in specific patients, we could further analyze and predict the progression of the disease. We adjusted the synthetic images by adding lesion patches with different shapes and intensities as in Fig. \ref{fig:editing} on a same patient. In this situation, image editing with UM-guided synthesis allows the user to gradually model the opacities, locations, and sizes of different types of lesions. In contrast, segmentation-guided synthesis failed to model the gradual change.  

\textbf{Lung Fibrosis Editing. }A UIP pattern \cite{raghu2018diagnosis} in the lung indicates a high risk of death and identifying such patterns in HRCT images is important. Because the population of UIP is small, it is helpful to generate UIP cases for classification tasks. We trained a synthesizer on UIP cases from the dataset used in \cite{walsh2018deep} and inferenced the UIP synthesizer with the unsupervised masks generated from non-UIP HRCT images. By doing so, we could successfully synthesize the UIP pattern on CT images from the normal controls and could produce more training samples for deep learning based UIP identification. In addition, our algorithms could potentially provide a longitudinal synthesis for UIP patients that might facilitate the prognosis of IPF. An example of the synthetic UIP pattern can be found in Fig. \ref{fig:uip}. 
\subsection{The Efficiency of Our UM-Guided Synthesis Compared  to DatasetGAN}
In our MICCAI submission, we used a pixel-wise classifier for synthetic image segmentation, similar to the datasetGAN \cite{zhang2021datasetgan}. However, the additional pixel-wise classifier consumes additional GPU memory for training and requires redundant annotation for synthetic images. In this paper, we proposed a semi-supervised based algorithm for segmentation. Compared to the pixel-wised classifier in the datasetGAN, we found that this semi-supervised strategy could not only reduce the additional GPU memory consumption but could also reduce the training time. The training time for 10 epochs of our semi-supervised algorithm (2 min, 10 labeled cases) is much faster compared to the pixel-wised classifier (32 min, 10 labeled cases) used previously \cite{xing2022cs}.

\section{Conclusions}
The goal of achieving better performance using fewer data with less human expert bias should be garnering more attention in medical image analysis. In this work, we proposed the UM-guided synthesis, which was guided by a set of unsupervised masks. These unsupervised masks not only yielded structural information but also provided intensity distribution guidance for synthetic images. Compared to the existing segmentation mask guided synthesis, our UM-guided synthesis can provide high-quality synthetic images with reasonable human anatomy and realistic lesions using only limited manual segmentation. More intriguingly, by validating using advanced image quality evaluation metrics, i.e., MM-FID and MM-STD, we demonstrated that our UM-guided synthesis could produce results with significantly higher fidelity, variety, and utility. \textcolor{black}{We demonstrate an accurate, generalizable and versatile UM-guided synthesis process which can have merit in the setting of improving medical diagnosis via automating accurate mask segmentation. The technical significance of our work is potentially the use of unsupervised masking towards efficiently conditioning the GAN-based synthesis process and can be applicable to other challenging problems of medical imaging research. Future directions are to investigate multiple physiology systems, and the theoretical exploration of mask-based GAN conditioning in enhancing further the synthesis process.}   

% \section{Acknowledgements}

\bibliographystyle{IEEEtran}
\bibliography{ref.bib}

\end{document}